\documentclass{sig-alternate-10pt}
\pdfoutput=1 

\usepackage{url}
\usepackage{graphicx}   % for EPS use the graphics package instead

\begin{document}
%
% --- Author Metadata here ---
%\conferenceinfo{AMC IMC}{2016 Santa Monica, California, USA}
%\CopyrightYear{2007} % Allows default copyright year (20XX) to be over-ridden - IF NEED BE.
%\crdata{0-12345-67-8/90/01}  % Allows default copyright data (0-89791-88-6/97/05) to be over-ridden - IF NEED BE.
% --- End of Author Metadata ---

\title{I Accidentally the Whole Internet}
\numberofauthors{3}
\author{
\alignauthor
Ivana Bachmann\\
\affaddr{NICLabs, Universidad de Chile}\\
\affaddr{Blanco Encalada 1975}\\
\affaddr{Santiago de Chile}\\
\email{ivana@niclabs.cl}
\alignauthor
Fernando Morales\\
\affaddr{NICLabs, Universidad de Chile}\\
\affaddr{Blanco Encalada 1975}\\
\affaddr{Santiago de Chile}\\
\email{fernando@niclabs.cl}
\and
Alonso Silva\\
\affaddr{Bell Labs, Nokia}\\
\affaddr{Centre de Villarceaux,}\\
\affaddr{Route de Villejust,}\\
\affaddr{91620 Nozay, France}\\
\email{alonso.silva@nokia.com}
\alignauthor
Javier Bustos-Jimenez\\
\affaddr{NICLabs, Universidad de Chile}\\
\affaddr{Blanco Encalada 1975}\\
\affaddr{Santiago de Chile}\\
\email{jbustos@niclabs.cl}
}
\additionalauthors{}

\maketitle
\begin{abstract}
Whether as telecommunications or power systems, networks are very important in everyday life. 
Maintaining these networks properly functional and connected, even under attacks or failures, is of special concern.

This topic has been previously studied with a whole network robustness perspective,
modeling networks as undirected graphs (such as roads or simply cables).
This perspective measures the average behavior of the network after its last node has failed.

In this article we propose two alternatives to well-known studies about the robustness of the backbone Internet:
to use a supply network model and metrics for its representation (we called it the Go-Index),
and to use robustness metrics that can be calculated while disconnections appear.

Our research question is: 
if a smart adversary has a limited number of strikes to attack the Internet,
how much will the damage be after each one in terms of network disconnection?  

Our findings suggest that in order to design robust networks
it might be better to have a complete view of the robustness evolution of the network,
from both the infrastructure and the users perspective.

\end{abstract}



\section{Introduction}

Transportation, electrical and telecommunication networks, to name a few, have become fundamental for the proper functioning of the modern world. For that reason it has become of extreme importance that these systems remain operative. However these systems are prone to failure due to malfunctions, catastrophes or attacks.

All these structures can be studied through complex networks by representing the components of the structure by nodes and interactions among the components by edges.

Since their correct functioning requires that the network is properly connected it is of great importance to study their ability to resist failures (either unintentional or targeted attacks). This ability is called robustness.

In this work, we focus on the scenario of targeted attacks by an adversary. We notice that this scenario corresponds to an upper bound on the damage of (unintentional or intentional) failures. 

We consider that an ``adversary'' should plan a greedy strategy aiming to maximize the damage with minimum number of strikes.
Thus, in this article we discuss the performance of attacks based on the edge betweenness centrality metric \cite{bersano2012metrics} over the Internet Backbone (the network formed by Internet exchange points, IXP), and its correlation with what users perceive from such networks if they want to receive content from the major content provider (Goo\-gle), with a metric called Go-Index.

We consider the Go-index measure with contains different supply network measures whose provider is Goo\-gle.
Just like economy uses the price of the Big Mac as a way of measuring purchasing power parity for its wide availability, 
here we measure the ability of the nodes to remain connected with Google for the same reason.

The idea to consider an IXP-based network as ``the Internet backbone'' is not new, previously has been used, for instance, as part of the ``internet core'' to study the inter-AS traffic patterns and an evolution of provider peering strategies \cite{labovitz2011internet}, for optimize the content delivery from Google via direct paths \cite{chiu2015we} (in this case filtering IXP ASs because they facilitate connectivity between peers), and the Internet Backbone Market \cite{buccirossi2005competition}. The importance of our study is based in with the use of the IXP network as a model for ``backbone Internet'', we can have a good approximation of Internet physical robustness.  As far as the authors knowledge, this is the first time that the robustness of IXP network is studied. 

The article is organized as follows, next section pre\-sents related work, followed by the methodology for building the IXP network, the attacking strategy using betweenness centrality and Go-Index (Sections \ref{bet} and \ref{go}). Conclusions are presented in Section \ref{conclusions}.

\section{Related work}
\label{relatedwork}

To study the robustness of a network its evolution against failure must be analyzed. On real world situations networks may confront random failures as well as targeted attacks.
For the latter, two main categories of attacking strategies have been defined: simultaneous and sequential attacks \cite{31holme2002attack}. Simultaneous attacks chose a set of nodes and remove them all at once while sequential attacks chose a node to remove and given the impact of this removal it chooses another node, proceeding in turns.

Targeted attacks have been thoroughly studied to analyze network robustness. Holme \textit{et al.} tested node degree and betweenness centrality strategies using simultaneous and sequential attacks. On \cite{33trajanovski2013robustness} simultaneous attacks based on network centrality measures and random attacks were studied.

The stability of scale-free network under degree-based attacks was studied on \cite{35yehezkel2012degree}. Experimental results are shown in \cite{iyer2013attack} who also consider sequential and simultaneous attacks as well as centrality measure strategies. To get closer to a real world strategy scenario on \cite{36gallos2006attack,37wu2007vulnerability} studied the resilience of scale-free networks to a variety of attacks with different amounts of information available to the attacker about the network.

On \cite{38booker2012effects} the impact of the effectiveness of the attack under observation error was studied. More recently \cite{ventresca2015network} studied sequential multi-strategy attacks using multiples robustness measures including the \textit{Unique Robustness Measure} ($R$-index)~\cite{schneider2011mitigation}.

The attacking strategies have been analyzed through the lens of an attacker, an adversary whose objective is to perform the most damage possible to the network. However the case of an adversary with a limited amount of strikes remain to be tested. Here this case is analyzed and an option to measure the robustness of a network in these circumstances is presented.

On \cite{32estrada2006network} was found that targeted attack can be more effective when they are directed to bottlenecks rather than hubs. On \cite{miuz} authors present partial values of $R$-index while nodes are disconnected, showing the importance of a well chosen robustness metric for performing the attacks.

The idea of planning a ``network attack'' using centrality measures has captured the attention of researchers and practitioners nowadays. For instance, Sterbenz et al.~\cite{sterbenz2011modelling} used bet\-ween\-ness-centrality (\textit{bcen}) for planning a network attack, calculating the \textit{bcen} value for all nodes, ordering nodes from higher to lower \textit{bcen}, and then attacking (discarding) those nodes in that order. They have shown that disconnecting only two of the top \textit{bcen}-ranked nodes, their packet-delivery ratio is reduced to $60\%$, which corresponds to~$20\%$ more damage than other attacks such as random links or nodes disconnections, tracked by link-centrality and by node degrees.  

A similar approach and results were presented by {\c{C}}etin\-kaya et al.~\cite{ccetinkaya2013modelling}. They show that after disconnecting only $10$ nodes in a network of $100+$ nodes the packet-delivery ratio is reduced to $0\%$. Another approach, presented as an improved network attack \cite{rak2010survivability, sydney2010characterising}, is to recalculate the betweenness-centrality after the removal of each node \cite{holme2002attack,molisz2006end}. They show a similar impact of non-recalculating strategies but discarding sometimes only half of the equivalent nodes. 

In the study of resilience after edge removing, Rosen\-kratz et al. \cite{rosenkrantz2009resilience} study backup communication paths for network services defining that a network is ``\textit{k-edge-failure-resilient if no matter which subset of k or fewer edges fails, each resulting subnetwork is self-sufficient}'' given that ``\textit{the edge resilience of a network is the largest integer k such that the network is k-edge-failure-resilient}''.

For a better understanding of network attacks and strategies, see  \cite{holme2002attack,molisz2006end,rak2010survivability, sydney2010characterising}. 

\section{Building the Internet backbone graph}
\label{graph}

\begin{figure}[ht!]
  \centering
  \includegraphics[width=\linewidth]{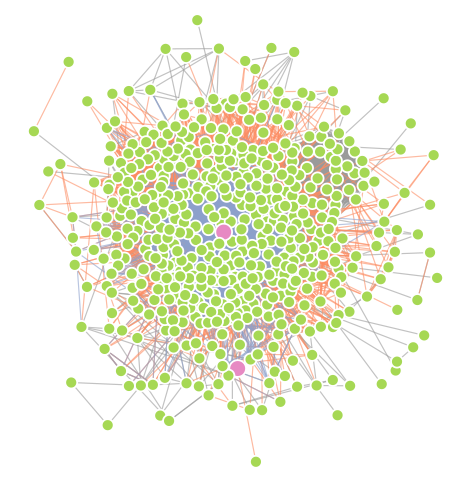}
  \caption{Peering Graph}
  \label{fig:graph}
\end{figure}

Internet peering is the contract (formal or informal) between two autonomous systems (AS) that agree to exchange traffic (and traffic routes) through a physical link. In \cite{Dhamdhere:2010:IFM:1921168.1921196} authors present that ``\textit{The core of the Internet is a multi-tier hierarchy of Transit Providers (TPs). About 10-20 tier-1 TPs, present in many geographical regions, are connected with a clique of peering links. Regional (tier-2) ISPs are customers of tier-1 TPs. Residential and small business access (tier-3) providers are typically customers of tier-2 TPs}''. Therefore, it is natural to think that the peering network is a coarse grained approximation of the Internet itself.  Thus, we used it to model Internet for studying its robustness.

From \url{peeringdb.com} we collected the autonomous systems (AS) from every Internet Exchange Point (IXP) and defined them as graph nodes. Therefore, an AS could belong to different IXPs and an IXP could have multiple ASs. Then, we connected the nodes if they fulfill at least one of the following rules:
\begin{itemize}
\item Physically linked ASs that exchange traffic 
\item ASs belonging to the same IXP
\item ASs belonging to the same facility
\end{itemize}
Where we considered IXP as public peering and facility as private peering.

\begin{figure}[ht!]
  \centering
  \includegraphics[width=\linewidth]{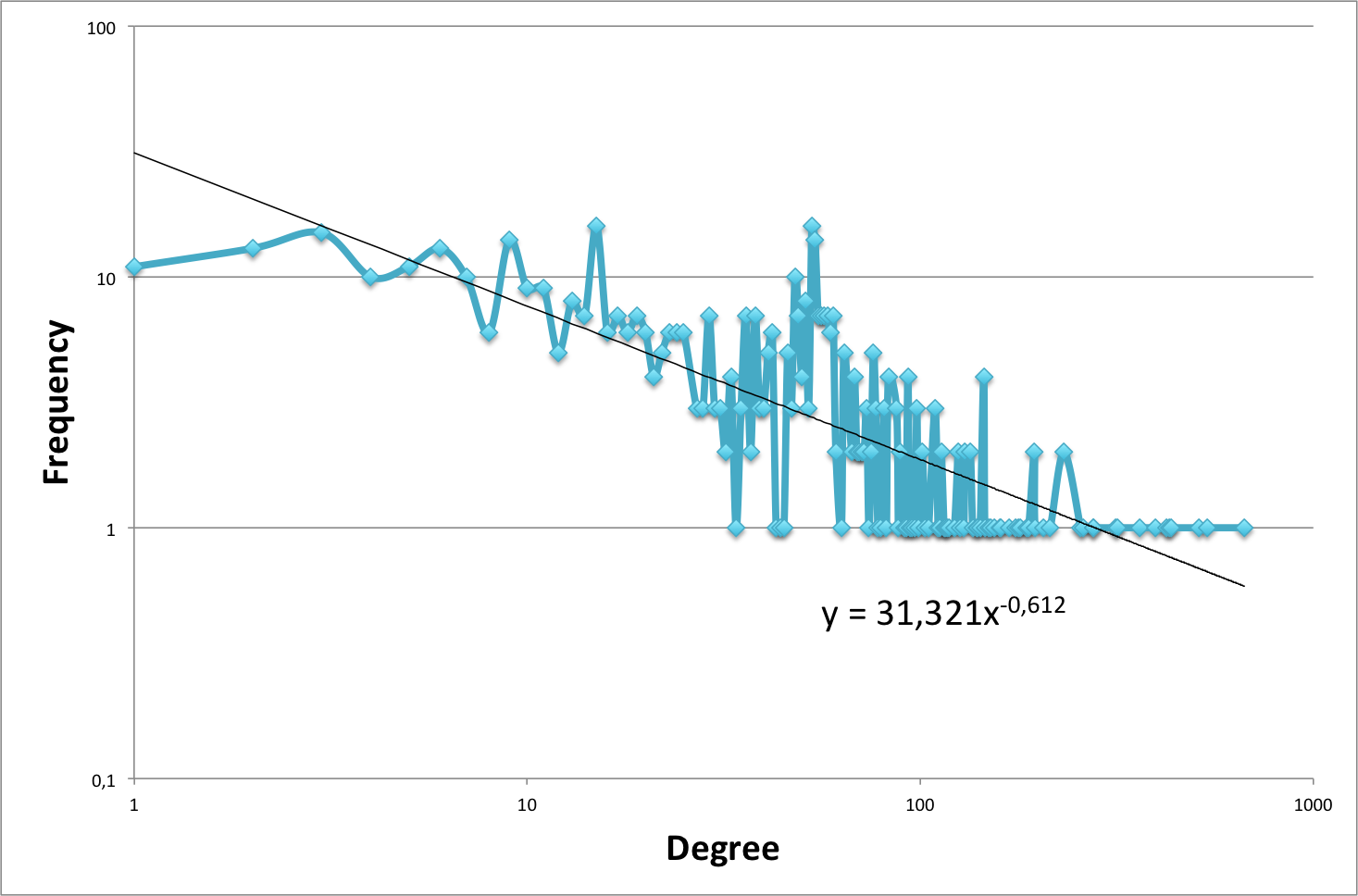}
  \caption{Degree distribution of IPX autonomous systems}
  \label{fig:degree}
\end{figure}

Figure~\ref{fig:graph} shows the resulting Graph, which has $522$ nodes and $14,294$ edges  (orange edges are public peering, blue are private peering, and green are direct network connection). The resultant network has a well connected core network with some isolated nodes at the edges. In Figure \ref{fig:degree} we present a degree distribution of nodes for our IPX Graph.

\section{Studying the robustness of the Internet backbone}
\label{bet}

Betweenness centrality is a metric that determines the importance of an edge by looking at the shortest paths between all of the pairs of remaining nodes. Betweenness has been studied as a resilience metric for the routing layer~\cite{smith2011network} and also as a robustness metric for complex networks \cite{iyer2013attack} and for internet autonomous systems networks~\cite{mahadevan2006internet} among others.  Betweenness centrality has been widely studied and standardized as a comparison base for robustness metrics, thus in this study it will be used for performance comparison. 

%Metrics used in other areas included the \textit{Effective Resistance} \cite{ellens2011effective} (robustness of electric power networks), \textit{Bridgeness} \cite{cheng2010bridgeness} (link importance based in its surrounded cliques), and some metrics for supply networks \cite{zhao2011achieving}.

If we plan a network attack by disconnecting edges with a given strategy, it is widely accepted to compare it against the use of  betweenness centrality metric, because the latter reflects the importance of an edge in the network \cite{iyer2013attack}. These attack strategies are compared by means of the \textit{Unique Robustness Measure} ($R$-index)~\cite{schneider2011mitigation}, defined as: 
\begin{equation}
R = \frac{1}{N}\sum_{Q=1}^{N} {s(Q)},
\end{equation}
where $N$ is the number of nodes in the network and $s(Q)$ is the fraction of nodes in the largest connected component after disconnecting edges using a given strategy.  Therefore, the higher the $R$-index, the better in terms of robustness.

Instead of just comparing the robustness, after the removal of all of the edges, we would like to study the behavior of the attacks after only a few strikes. To do so, we define a variant of the $R$-index which takes into account only the first $n$ strikes of an attack. Thus, for a simultaneous attack (where the nodes are ranked by a metric only once at the beginning), the $R_n$-index is defined as:
\begin{equation}
R_n = \frac{1}{n}\sum_{Q=1}^{n} {s(Q)}.
\end{equation}
That is, the area under the curve produced by the largest connected component ratio (compared to the whole network) until $n$.

For a sequential attack, the order of node disconnection is recomputed after each disconnection. Similar to the $R$-index, notice that the lower the $R_n$-index, the more effective the attack is, since that gives us a higher reduction of robustness. 

\begin{figure}[th!]
  \centering
  \includegraphics[width=\linewidth]{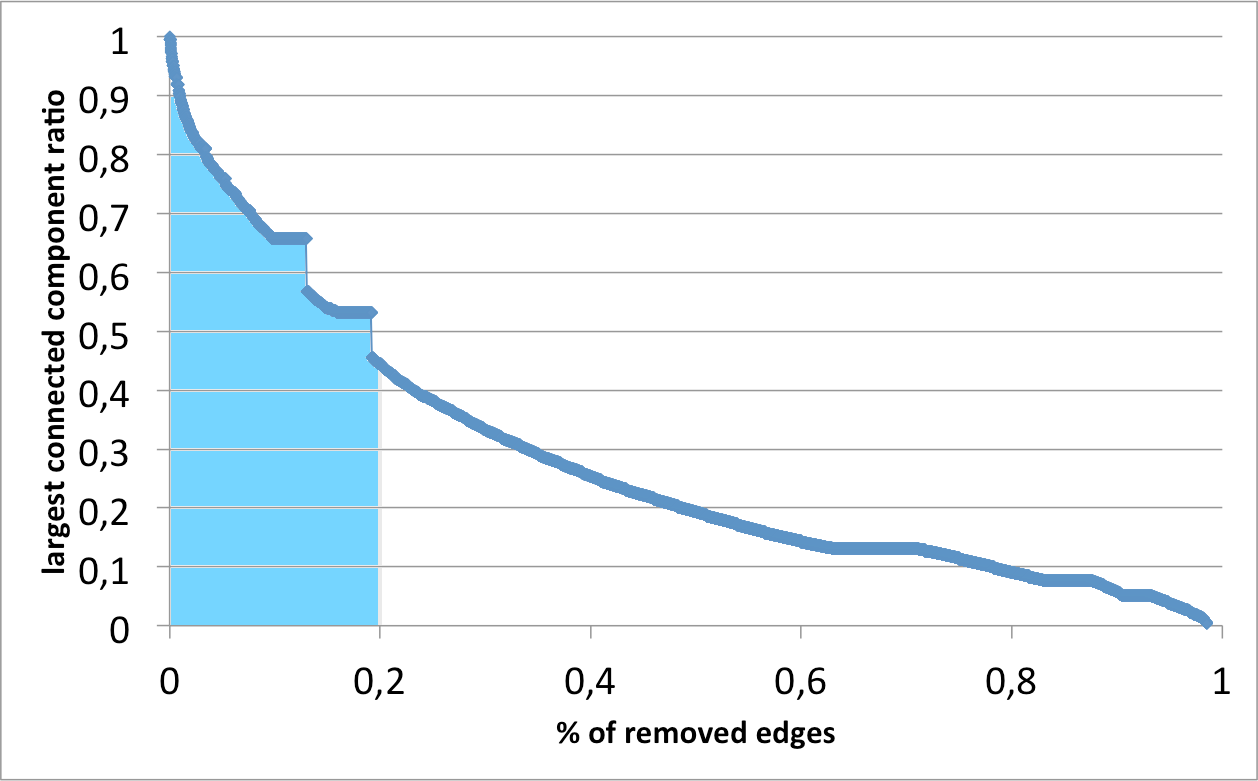}
  \caption{\% of the largest component size compared to the original network. In the plot, $R_{20\%}$-index is marked in cyan.}
  \label{fig:bet-attack}
\end{figure}

Results are shown in Figure \ref{fig:bet-attack}. We tested sequential attacks: At each strike, the next edge to disconnect was the one with the highest betweenness value.  The figure shows the behavior of the $R_n$-index in our IPX Graph. The strategy proves to be very effective in attacks, disconnecting half of the network removing only 20\% of the edges, more than 30\% of the nodes after removing 10\% of edges, and 10\% of nodes after 1\% of edges.

\section{The Go-Index: If you cannot see Google, you are not connected}
\label{go}

Google has been reported as having the 40\% of Internet traffic in 2013\footnote{See the Forbes article at \url{http://goo.gl/aHdeiN}}. Thus, given the Google peering policies\footnote{See \url{https://peering.google.com/#/options/peering}.} and knowing the Google policies to interconnect their datacenters \cite{jain2013b4}, we can also study Internet as an information supply network, adapting to Internet the metrics presented in \cite{zhao2011achieving}:
\begin{enumerate}
\item \textbf{Supply Availability} (SAR): The percentage of ASs that have access to Google from at least one of its ASs.
\item \textbf{Network Connectivity} (NetCON): The number of ASs in the largest functional sub-network, in which there is a path between any pair of ASs and there exist at least one of the Google ASs.
\item \textbf{Best Delivery Efficiency} (BDE): The reciprocal of the average of each demand AS's shortest supply path length to its nearest Google AS.  Values go from $1$ (everyone is connected directly with a Google AS) to $0$ (there are only Google ASs in the network).
\item \textbf{Average Delivery Efficiency} (ADE): The average inverse supply path length for all possible \{Network AS,Google AS\} pairs, adjusted by a weighting factor for each path (in our study all Google ASs have the same importance). In this case, values go from $2$ (everyone is connected to both Google ASs directly) to $0$ (nobody is connected with Google ASs).
\end{enumerate}
Notice that Google delegated some services at ISPs autonomous systems \cite{calder2013mapping}, nevertheless they must eventually connect with  Google backbone for updating.  We called the tuple \{1,2,3,4\} the \textit{Go-Index}, that is, the supply network measures whose provider is Google. 

\begin{figure}[th!]
  \centering
  \includegraphics[width=\linewidth]{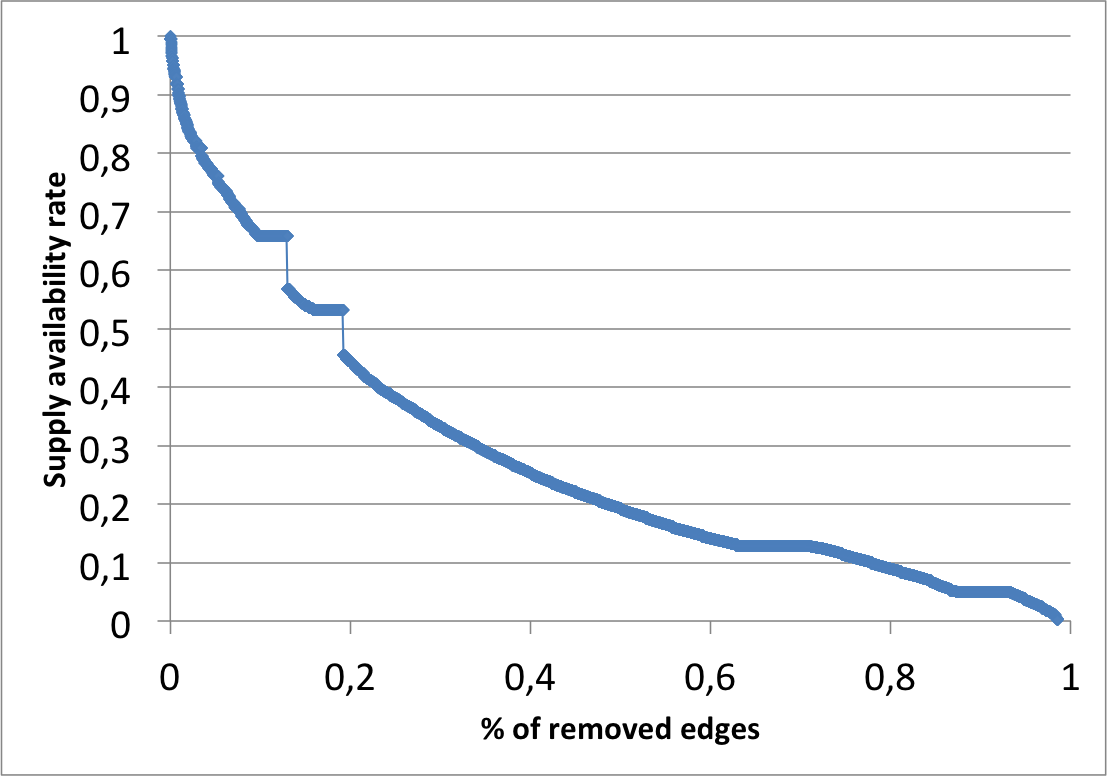}
  \caption{Supply Availability Ratio}
  \label{fig:sar}
\end{figure}

\begin{figure}[th!]
  \centering
  \includegraphics[width=\linewidth]{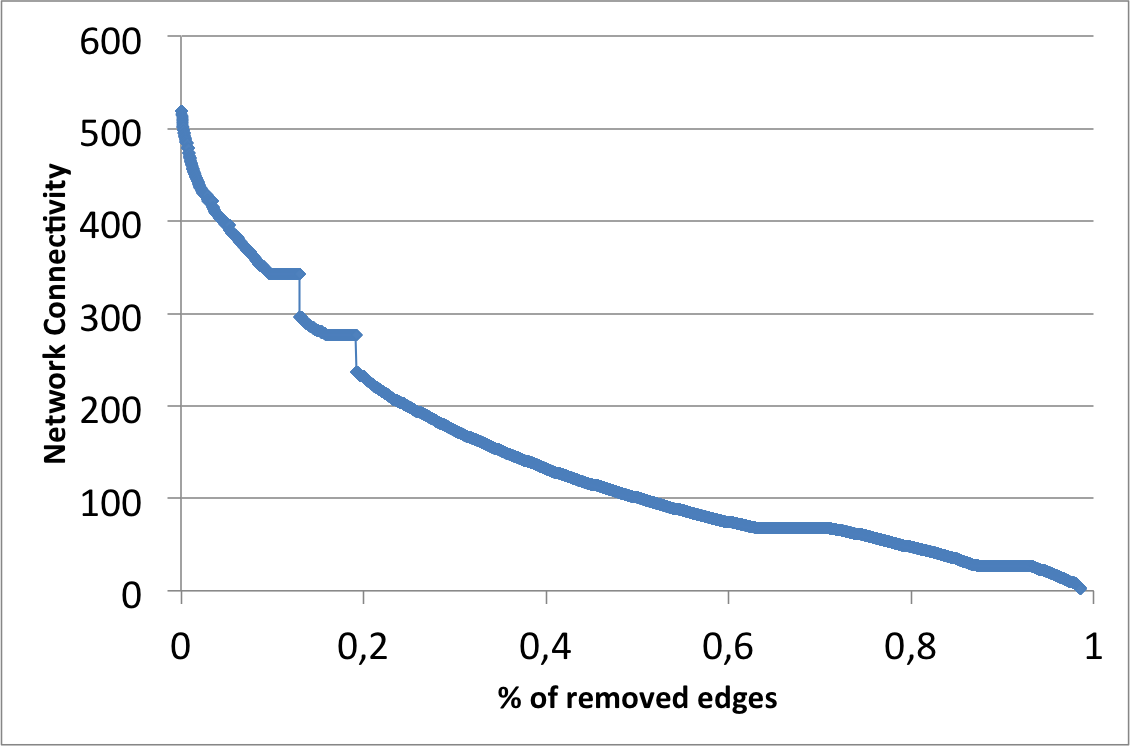}
  \caption{Network Connectivity}
  \label{fig:netcon}
\end{figure}

\begin{figure}[th!]
  \centering
  \includegraphics[width=\linewidth]{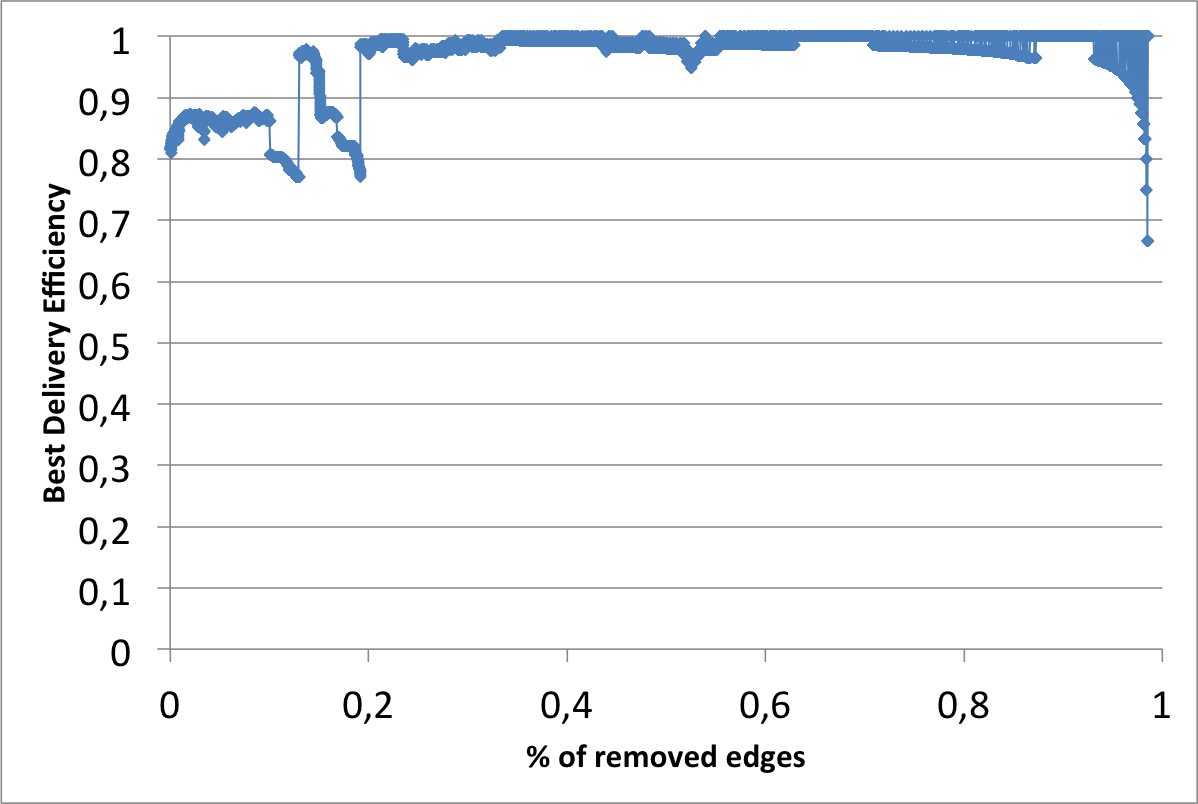}
  \caption{Best Delivery Efficiency}
  \label{fig:bde}
\end{figure}

\begin{figure}[th!]
  \centering
  \includegraphics[width=\linewidth]{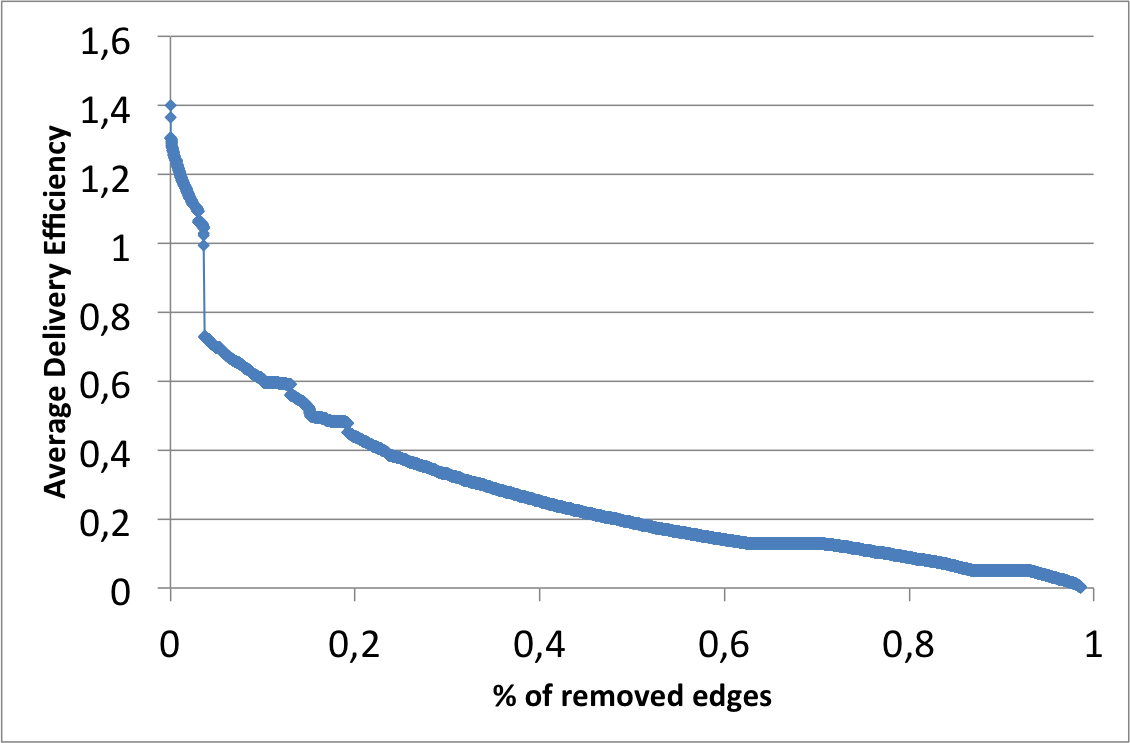}
  \caption{Average Delivery Efficiency}
  \label{fig:ade}
\end{figure}

Using the same attack strategy from previous section, we calculated the Go-Index after edge removal (removing the edge with higher betweenness). The results are presented in Figure \ref{fig:sar} (Supply Availability), \ref{fig:netcon} (Network Connectivity), \ref{fig:bde} (Best Delivery Efficiency), and \ref{fig:ade} (Average Delivery Efficiency).  

The first two metrics are very related with the largest connected component ratio, which in this study include at least one of the two Google ASs, that is, AS15169 and AS36040 (marked in pink at Figure \ref{fig:graph}, the former in the center and the latter in the edge of the network).  Therefore, no new information are provided by those metrics.

The following two metrics are based  in ``how far is Google from a given autonomous system'', but  BDE considers only the connected component that includes Google ASs, by itself it has no information about the isolated portion of the network that cannot reach the Google ASs, therefore BDE improves when large subnets are disconnected from the core network that contains the Google ASs.  

Note that for a user inside that core network the main content provider always exists and there are no indications that the network is falling apart (or losing half of its members, as produced by eliminating 20\% of its edges), this can be appreciated through BDE. Nevertheless, the big picture is different: after having only the 5\% of the network disconnected one of Google ASs is isolated, showing that from this point the supply network is only maintained by AS15169. The perception error is corrected when ADE is used because it includes all nodes in its calculus.

Then, the Go-Index accomplish its objectives, reflecting both infrastructure (SAR+NetCON+ADE) and user perception (BDE+ADE), for Internet robustness studies.

\section{Conclusions and Future Work}
\label{conclusions}

In this paper we have presented how robust the Internet backbone (the peering AS network) would be
if an adversary can choose wisely which physical link he will cut (or if a very unlucky accident happen).
Following the recommendations, the chosen one would be the edge with higher
betweenness centrality value. 

Using this strategy the adversary is capable of disconnecting half of the network by removing only 20\% of the edges, more than 30\% of the nodes after removing 10\% of the edges, and 10\% of nodes after 1\% of the edges as we have seen with the values of $R_{n}$ index in Section  \ref{bet}.

Furthermore, we consider the Internet as a (information) supply network,
and considering that Google is the main Internet content provider, 
and propose to study the Internet backbone with the Go-Index 
(the adaptation of metrics presented in \cite{zhao2011achieving}  for supply networks).

The metrics used were Supply Availability, Network Connectivity (both highly correlated with the larger connected component ratio), 
Best Delivery Efficiency, and Average Delivery Efficiency.

If only Best Delivery Efficiency is considered, the network can be declared robust 
because a user located inside the core network, which is always connected with a Google AS, 
will not perceive that the network is being disconnected. The Go-index will correct that perception since it also contains the Average Delivery Efficiency which includes all nodes in its calculus. 

As future work we plan to apply similar studies to other Internet infrastructures, 
such as country-based fiber interconnection, submarine Internet cables, etc. 
Also, we plan to improve the metrics for robustness reflecting both 
the infrastructure part (such as Rn index) and the user perception 
(such as Best Delivery Efficiency/Average Delivery Efficiency).

\end{document}